\documentclass[prb,twocolumn,showpacs,preprintnumbers,amsmath,amssymb]{revtex4}
\usepackage[dvips]{graphicx}
\usepackage{dcolumn}
\usepackage{bm}
\usepackage{times}
\usepackage{mathptmx}
\usepackage{color}
\usepackage{pifont}
\usepackage[latin1]{inputenc}

\begin{document}

\title{Second harmonic generation on incommensurate structures:\\ The case of multiferroic MnWO$_4$}

\author{D. Meier}
\author{N. Leo}
\author{G. Yuan}
\author{Th.\ Lottermoser}
\author{M. Fiebig}
\affiliation{%
Helmholtz-Institut f\"ur Strahlen- und Kernphysik, Universit\"{a}t Bonn\\
Nussallee 14--16, 53115 Bonn, Germany
}%

\author{P. Becker}
\author{L. Bohat\'y}
\affiliation{
Institut f\"ur Kristallographie, Universit\"at zu K\"oln\\
Z\"ulpicher Strasse 49b, 50674 Köln, Germany
}%

\date{\today}

\begin{abstract}
A comprehensive analysis of optical second harmonic generation (SHG) on an incommensurate (IC) magnetically ordered
state is presented using multiferroic MnWO$_4$ as model compound. Two fundamentally different SHG
contributions coupling to the primary IC magnetic order or to secondary
commensurate projections of the IC state, respectively, are distinguished. Whereas the latter can
be described within the formalism of the 122 commensurate magnetic point groups the former
involves a breakdown of the conventional macroscopic symmetry analysis because of its sensitivity
to the lower symmetry of the {\it local environment} in a crystal lattice. Our analysis thus
foreshadows the fusion of the hitherto disjunct fields of nonlinear optics and IC order in
condensed-matter systems.
\end{abstract}


\maketitle

\section{Introduction: imaging incommensurate order}

A crystal structure is called incommensurate (IC), if a periodic deviation of atomic parameters
from the basic crystal structure cannot  be related to the periodicity of the underlying crystal structure
by a rational number.\cite{Steurer08a} The violation of the translation symmetry of the
basic crystal structure by the IC long-range order manifests in an
astonishing diversity of unusual physical effects so that systems with IC phases attracted a lot
of attention since Dehlinger reported the first aperiodic crystal in 1927.\cite{Dehlinger27a}
Nowadays, IC order is known to play an essential role in quasicrystals\cite{Shechtman84a}, liquid
crystals,\cite{Fontes88a} magnetic multilayers,\cite{Salamon86a} or polymers.\cite{Taylor86a}
Other examples for the omnipresence of IC phases can be found among systems with strong electronic
correlations such as high-temperature superconductors,\cite{Wells97a} colossal-magnetoresistance
manganites,\cite{Chen96a} or multiferroics.\cite{Harris05a} The pronounced reduction of symmetry
by the IC order reflects the complexity of the underlying microscopic interactions.
Macroscopically, this can lead to a variety of phases that become allowed in the IC state
\cite{Toudic08a} and enrich the phase diagram of the host material
--- a key ingredient for the design of multifunctional materials.

Despite the longstanding interdisciplinary interest in IC structures, their theoretical and
experimental analysis is still a challenge. For instance, it was realized only recently that an IC
magnetic spiral may induce a spontaneous electric polarization, and a controversial discussion
about the microscopic mechanisms of this coupling is persisting.\cite{Katsura05a,Mostovoy06a,Sergienko06a} For analyzing the
order parameters and domain structures of an IC structure, a group-theoretical approach based on
Landau theory of phase transitions is usually chosen. A
description of the IC state in a superspace with $>3$ dimensions\cite{Wolff74a} or, alternatively,
by considering the irreducible representations of the three-dimensional space group in the
presence of a modulated structure\cite{Toledano87a} are possible. Experimentally, neutron and
X-ray diffraction are usually employed for investigating aperiodic spin, charge, or lattice
modulations. Yet, although these established techniques constitute a versatile tool for probing IC
structures, essential ingredients characterizing the IC state remain inaccessible. In particular,
this applies to the domains associated to the IC state. On the one hand, the spatial resolution of
a diffraction experiment is often limited because of the integrating nature of the
experiment. On the other hand, most of the IC magnetic structures do not carry a macroscopic
magnetization so that it is difficult to couple to the order parameter and its spatial
distribution. However, since domains and domain walls determine many macroscopic physical
properties and, thus, the technological feasibility of a compound, access to these domains in a
convenient way is highly desirable, particularly because it is expected that the IC nature of the
ordered state will lead to domains with novel properties.

A technique with inherent spatial resolution and access even to ``hidden'' ordered structures is
optical second harmonic generation (SHG).\cite{Fiebig05a} For decades SHG has been applied to
image structural, magnetic, or electric domains and to reveal spatial correlation effects between
them.\cite{Rasing97a,Uesu95a} More recently, SHG was even used to resolve the sub-picosecond
dynamics of a magnetic order parameter.\cite{Regensburger00,Fiebig08a} However, almost any form of
long-range order probed by SHG thus far was commensurate. In the very small number of
investigations on IC structures it was demonstrated that the emergence of the IC state is
reflected by changes in the SHG yield, but a theoretical analysis of the detected signal was never
attempted.\cite{Lemanov91a,Ibragimov02a,Qin98a,Golovko80a} The same holds for recent
investigations on IC magnetic ferroelectrics where SHG was used for imaging magnetic and
ferroelectric domains without performing a detailed discussion of the SHG contributions and the IC
symmetries involved.\cite{Meier09a,Meier09c} Thus, a resilient framework categorizing the
nonlinear interactions between the electromagnetic light fields and an aperiodically ordered state
of matter is yet to be developed.

In this report we present a comprehensive analysis of optical SHG on an IC magnetically ordered
state. Using multiferroic MnWO$_4$ as model compound, two principally different SHG contributions
coupling to the symmetry-breaking order parameters are identified. The first one corresponds to a
commensurate projection of the IC magnetic state that is represented by an improper order
parameter and can be described within the 122 commensurate magnetic point groups. The second one
reflects the entire IC structure of the compound and involves a breakdown of the conventional
macroscopic symmetry analysis since local symmetries become involved. With our analysis we
foreshadow the fusion of the hitherto disjunct fields of nonlinear optics and IC order in
condensed-matter systems.

\section{Incommensurate order in multiferroic M\symbol{110}WO$_4$}

Multiferroics are compounds uniting at least two primary forms of (anti-) ferroic order in the
same phase of a material. The majority of research activities is devoted to the magnetically
ordered ferroelectrics since the coexistence of magnetic and electric order is a potential source
for magnetoelectric cross-correlations allowing one to manipulate magnetic order by electric
fields (or vice versa). The most pronounced, so-called ``gigantic'', magnetoelectric coupling
effects are observed when IC magnetic order promotes the emergence of a spontaneous electric
polarization that is rigidly coupled to the magnetic order parameter. Because of the novel and
unusual nature of this manifestation of IC order and the intense desire to advance our
understanding of it, the choice of an IC magnetically induced ferroelectric as model system for
unravelling the relation between IC order and SHG suggests itself.

We select MnWO$_4$ because it combines a relatively simple crystallograhic structure with a
variety of IC phases. The magnetic phase diagram is well established and the general feasibility
of an approach by SHG has already been demonstrated\cite{Meier09a,Meier09c,Meier09b} so that we
can now focus on analyzing the correlation between SHG and the IC state. MnWO$_4$ is closely
related to compounds such as TbMnO$_3$, Ni$_2$V$_3$O$_8$, or
CuO\cite{Kenzelmann05a,Kimura08a,Lawes05a} so that the results gained here will be of general use.

\begin{figure}
\centering
\includegraphics[width=0.48\textwidth]{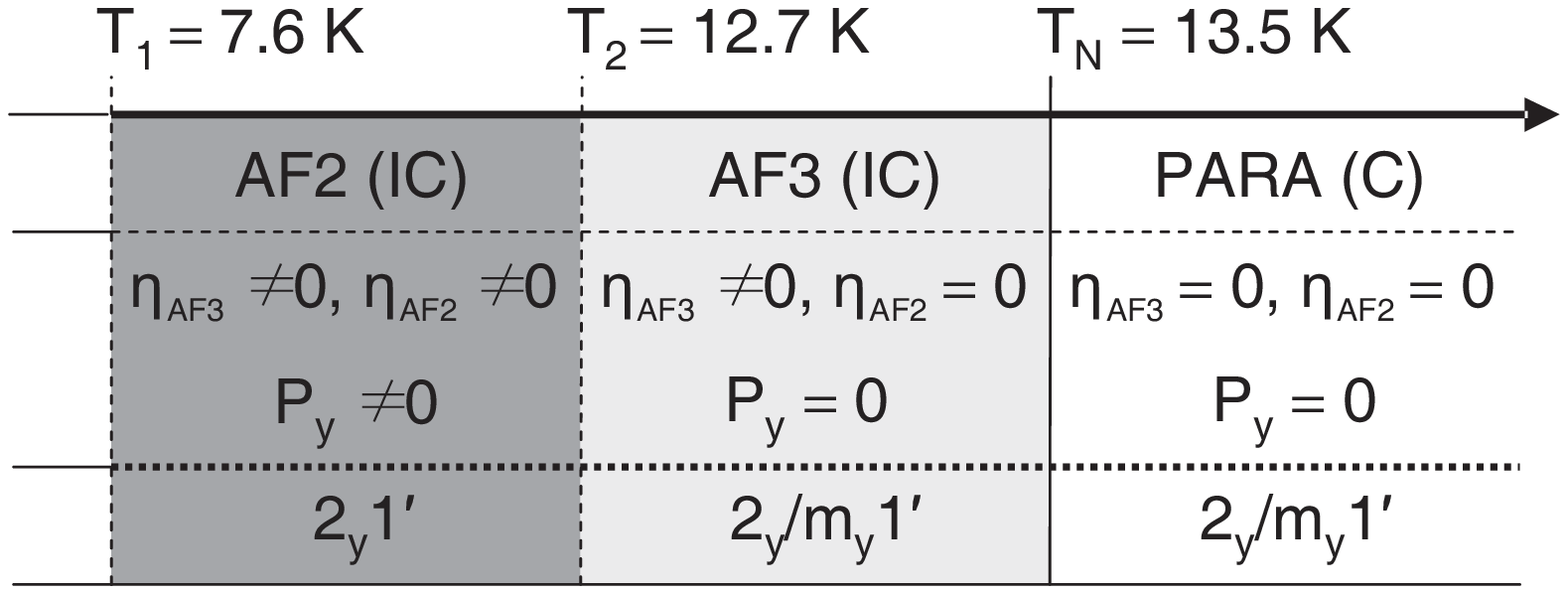}
\caption{\label{fig:fig1} Overview of the magnetic phases\cite{Arkenbout06,Taniguchi06a}, the
symmetry-breaking order parameters\cite{Toledano10a,Harris07a}, and the corresponding point groups
for multiferroic MnWO$_4$. The commensurate AF1 phase at $T<7.6~K$ is not discussed in this
manuscript and therefore not shown.}
\end{figure}

The multiferroic phase of MnWO$_4$ is governed by two fundamentally different types of order
parameters: (i) a set of two primary magnetic order parameters ($\eta_{\text{\tiny{AF3}}}$,
$\eta_{\text{\tiny{AF2}}}$), and (ii) one secondary electric order parameter ($P_y^{\rm sp}$). The
sequence of phase transitions leading to the two classes of order parameters in MnWO$_4$ is
sketched in Fig.~\ref{fig:fig1}.
Below $T_{\rm N}=13.5$~K the magnetic moments of Mn$^{2+}$ align in a
collinear way along the easy axis while their magnitudes are sinusoidally modulated with $\mathbf
k_{\text{\tiny{AF3}}}=(-0.214,\frac{1}{2},0.457)$~\cite{Lautenschlaeger93a}. The resulting IC
spin-density wave breaks the translation symmetry of the lattice but conserves the inversion
symmetry of the crystal. This so-called AF3 phase can be described by a single two-dimensional
magnetic order parameter
\mbox{$\eta_{\text{\tiny{AF3}}}=(\sigma_{\text{\tiny{AF3}}}e^{i\theta_{\text{\tiny{AF3}}}},
\sigma_{\text{\tiny{AF3}}}e^{-i\theta_{\text{\tiny{AF3}}}})$}~\cite{Toledano10a,Harris07a}. Upon
further cooling an additional transverse spin component orders at $T_2=12.7$~K. In this ${\rm
AF3}\to{\rm AF2}$ transition the spin-density wave turns into an elliptical spin spiral without
changing the IC periodicity of the modulation so that ${\mathbf k}_{\text{\tiny{AF2}}}={\mathbf
k}_{\text{\tiny{AF3}}}$. The ${\rm AF3}\to{\rm AF2}$ transition is driven by a second magnetic
order parameter
$\eta_{\text{\tiny{AF2}}}=(\sigma_{\text{\tiny{AF2}}}e^{i\theta_{\text{\tiny{AF2}}}},
\sigma_{\text{\tiny{AF2}}}e^{-i\theta_{\text{\tiny{AF2}}}})$. The coexistence of
$\eta_{\text{\tiny{AF2}}}$ and $\eta_{\text{\tiny{AF3}}}$ breaks the inversion symmetry and
induces a spontaneous polarization $P_y^{\rm sp}$ along the $y$ axis ($y\,\|\,b$) according
to~\cite{Toledano10a,Harris07a,footnote1}
\begin{equation}\label{eq:indP}
  P_y^{\rm sp} \propto \sigma_{\text{\tiny{AF2}}}\sigma_{\text{\tiny{AF3}}} \; ,
\end{equation}
thus constituting the multiferroic phase of MnWO$_4$. The correlation expressed by
equation~\eqref{eq:indP} implies that the IC nature of the primary (proper) order parameters
$\eta_i$ with amplitudes $\sigma_i$ is projected onto the secondary (improper; here:
pseudo-proper\cite{Toledano10a}) order parameter $P_y^{\rm sp}$. The projection corresponds to a
commensurate net polarization that may be additionally modulated in an IC way around its net
value.\cite{Toledano10a} Note that the incommensurability of $P_y^{\rm sp}$ is not mandatory and,
even if present, remains mostly unnoticed because experimental techniques probing the spontaneous
polarization such as pyroelectric measurements measure the commensurate contribution only.
Macroscopically the polarized state can be therefore described by the point-group formalism used
for commensurate structures, as in the lower part of Fig.~\ref{fig:fig1}.

\section{Optical second harmonic generation}

An experimental method that is known to be particularly sensitive to the point symmetry of a
crystal is optical SHG\cite{Fiebig05a}. It is described by the equation\cite{Boyd08}
\begin{equation}\label{eq:SHG}
  S_i(2\omega) = \epsilon_0 \chi_{ijk} E_j(\omega) E_k(\omega)  \; .
\end{equation}
An electromagnetic light field $\mathbf{E}$ at frequency $\omega$ is incident on a crystal,
driving a charge oscillation $\mathbf{S}(2\omega)$, which acts as source of a frequency-doubled
light wave of the intensity $I_{\rm SHG}\propto|\mathbf{S}(2\omega)|^2$. The source term
$\mathbf{S}$ can be written as multipole expansion
\begin{equation}\label{eq:sourceterm}
  \mathbf{S}(2\omega)=\mu_0\frac{\partial^2\mathbf{P}(2\omega)}{\partial t^2}
  +\mu_0\left({\mathbf{\mathbf\nabla}}\times\frac{\partial\mathbf{M}(2\omega)}{\partial t}\right)
  -\mu_0\left(\frac{\partial^2({\mathbf{\mathbf\nabla}}\hat{Q})}{\partial t^2}\right)  \; .
\end{equation}
with $\mathbf{P}(2\omega)$, $\mathbf{M}(2\omega)$, and ${\mathbf{\mathbf\nabla}}\hat{Q}(2\omega)$
as electric-dipole (ED), magnetic-dipole (MD), and electric-quadrupole (EQ) contribution,
respectively. The nonlinear susceptibility $\chi_{ijk}$ couples incident light fields with
polarizations $j$ and $k$ to a SHG wave with polarization $i$. According to Neumann's principle
the symmetry of a compound determines the set of nonzero components
$\chi_{ijk}$.\cite{Birss66,Fiebig05a} As a consequence, ED-SHG vanishes in centrosymmetric media
while MD-SHG and EQ-SHG remain allowed.

Using Eqs.~(\ref{eq:SHG}) and (\ref{eq:sourceterm}) and applying symmetry-dependent selection
rules\cite{Birss66} the tensor components summarized in Table~\ref{tab:chiMnWO} are derived as
contributions to SHG in MnWO$_4$.
\begin{table}
\centering
\begin{tabular}{lcll}
\hline \hline
Point  &Incident    &MD/EQ-SHG      &ED-SHG\\
group  & light      &contributions  &contributions\\
\hline
$2_y/m_y1'$ & $k\,\|\,x$  &$\chi^{\text{\tiny{MD,EQ}}}_{yyy}$, $\chi^{\text{\tiny{MD,EQ}}}_{yzz}$, $\chi^{\text{\tiny{MD,EQ}}}_{zyz}$& ---\\
 (AF3) & $k\,\|\,y$  & --- & --- \\
  & $k\,\|\,z$  &$\chi^{\text{\tiny{MD,EQ}}}_{yyy}$, $\chi^{\text{\tiny{MD,EQ}}}_{yxx}$, $\chi^{\text{\tiny{MD,EQ}}}_{xyx}$ & ---\\
\hline
$2_y1'$ & $k\,\|\,x$  &$\chi^{\text{\tiny{MD,EQ}}}_{yyy}$, $\chi^{\text{\tiny{MD,EQ}}}_{yzz}$, $\chi^{\text{\tiny{MD,EQ}}}_{zyz}$ & $\chi^{\text{\tiny{ED}}}_{yyy}$, $\chi^{\text{\tiny{ED}}}_{yzz}$, $\chi^{\text{\tiny{ED}}}_{zyz}$\\
 (AF2) & $k\,\|\,y$  & --- & ---\\
  & $k\,\|\,z$  &$\chi^{\text{\tiny{MD,EQ}}}_{yyy}$, $\chi^{\text{\tiny{MD,EQ}}}_{yxx}$, $\chi^{\text{\tiny{MD,EQ}}}_{xyx}$ & $\chi^{\text{\tiny{ED}}}_{yyy}$, $\chi^{\text{\tiny{ED}}}_{yxx}$, $\chi^{\text{\tiny{ED}}}_{xyx}$ \\
\hline \hline
\end{tabular}
\caption{\label{tab:chiMnWO} Non-zero SHG tensor components $\chi_{ijk}$ allowed on the basis of a
symmetry analysis within the 122 magneto-crystalline point groups.\cite{Birss66} The SHG tensor
is related to a Cartesian reference system ($x$, $y$, $z$) as defined in~\onlinecite{footnote1}. Calculations were
done for the magnetic AF3 phase and the multiferroic AF2 phase of MnWO$_4$ for light propagating
along the $x$, $y$, and $z$ axis. Note that SHG contributions with $i\neq j\neq k$ cannot be
detected with light propagating along $x$, $y$, or $z$ so that they are omitted. MD-SHG and
EQ-SHG lead to the same selection rules and are therefore not distinguished.}
\end{table}
Note that Table~\ref{tab:chiMnWO} is based on the use of point
groups (as in Fig.~\ref{fig:fig1}). This is based on the assumption that translation operations
and translation symmetries can be neglected in optical experiments since they are not resolved by the light.
More quantitatively, the variation of the electromagnetic amplitude of the light field across the
expansion of a unit cell (as limit of the non-primitive lattice translations) is considered to be
small enough to be neglected. With respect to Table~\ref{tab:chiMnWO} this means that the
aperiodic nature of the order parameters with their violation of the translation symmetry is also
not taken into account. The corresponding SHG light will be termed commensurate (C-) SHG in the
following in order to emphasize that averaged, lattice-periodic properties are considered.
Table~\ref{tab:chiMnWO} reveals that for light fields propagating along the $y$ axis ($k\,\|\,y$)
SHG is forbidden in all phases. For $k\,\|\,x$ or $k\,\|\,z$ MD-SHG and EQ-SHG can occur at any
temperature whereas ED-SHG is restricted to the multiferroic AF2 phase.

A detailed discussion of the technical aspects of SHG in ferroic systems including the
experimental setup used here is found in Refs.\ \onlinecite{Fiebig05a} and \onlinecite{Meier09a}.
In the present experiment, melt-grown MnWO$_4$ single crystals\cite{Becker07a} were processed into sets of
polished platelets with a thickness of about 100~$\mu$m with each of the samples being oriented
perpendicular to the one of the Cartesian axes $x$, $y$, $z$.\cite{footnote1} The samples were illuminated at normal incidence
in a transmission setup by light pulses of $2-5$~mJ and $3-8$~ns at a repetition rate of
$10-40$~Hz.

\section{Experimental results and discussion}

Based on Table~\ref{tab:chiMnWO} the experiments will proceed as follows: \textbf{A.} identification of
electronic transitions in the optical range and of any SHG contributions not coupling to magnetic
or electric long-range order; \textbf{B.} identification and analysis of C-SHG contributions; \textbf{C.}
identification and analysis of IC-SHG contributions.

\subsection{Optical transitions and paramagnetic SHG contributions}

\begin{figure}
\centering
\includegraphics[width=0.40\textwidth]{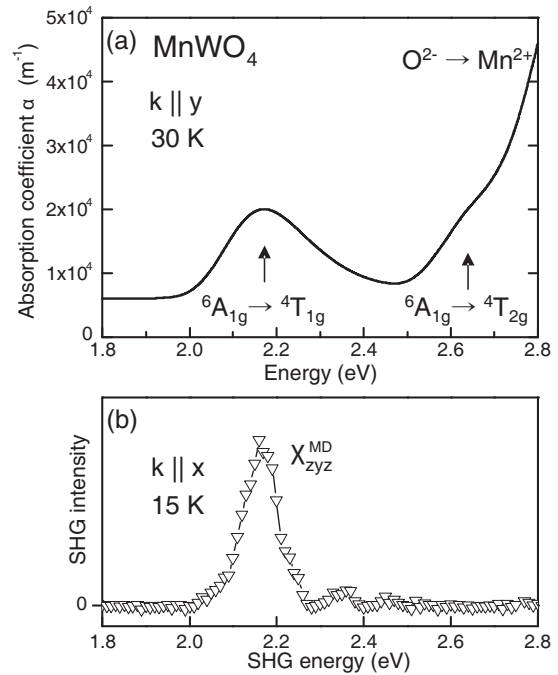}
\caption{\label{fig:fig2} Optical transitions and paramagnetic SHG contributions in MnWO$_4$.
(a) Linear absorption spectrum of MnWO$_4$ obtained with $x$-polarized light propagating along the
$y$ axis of the crystal. At low temperature two intra-atomic and one charge-transfer transitions
are identified and labelled accordingly. (b) SHG spectra obtained in the paramagnetic phase with
light incident along the $x$ axis. Only crystallographic SHG from
$\chi_{zyz}^{\text{\tiny{MD,EQ}}}$ is observed.}
\end{figure}

For identifying the electronic transitions of the Mn$^{2+}$ ion and the crystallographic
background contributions to SHG, Fig.~\ref{fig:fig2} shows linear absorption spectra and
polarization-dependent SHG spectra in the paramagnetic phase of MnWO$_4$.
The absorption spectrum
taken with $x$-polarized light incident along the $y$ axis exhibits a steep increase beyond 2.7~eV
which corresponds to the lowest O$^{2-}$--Mn$^{2+}$ charge transfer.\cite{Nogami08a,Taniguchi10a}
In addition, absorption peaks with a width in the order of 0.1~eV are observed and marked by black
arrows in Fig.~\ref{fig:fig2}(a). These peaks at 2.18~eV and 2.65~eV are assigned to the
intra-atomic \mbox{${}^6A_{1g}\rightarrow{}^4T_{1g}$} and \mbox{${}^6A_{1g}\rightarrow{}^4T_{2g}$}
transitions between the Mn$^{2+}(3d^5)$ orbitals.\cite{Huffmann69a}

In agreement with the symmetry analysis ED-SHG contributions were not observed above $T_{\rm N}$.
The signal in Fig.~\ref{fig:fig2}(b) shows the \mbox{${}^6A_{1g}\rightarrow{}^4T_{1g}$} transition
and is attributed to the $\chi^{\text{\tiny{MD,EQ}}}_{zyz}$ component, {\it i.e.}, to a
crystallographic C-SHG signal according to Table~\ref{tab:chiMnWO}. Note that out of the six
components $\chi^{\text{\tiny{MD,EQ}}}_{ijk}$ that are allowed $\chi^{\text{\tiny{MD,EQ}}}_{zyz}$
is the only one that was actually observed. Its occurrence is limited to the (100) sample where it
was suppressed henceforth by setting the polarization of the incident light along $y$ or $z$, but
not in between.

\subsection{Commensurate SHG contributions}

\begin{figure}
\centering
\includegraphics[width=0.40\textwidth]{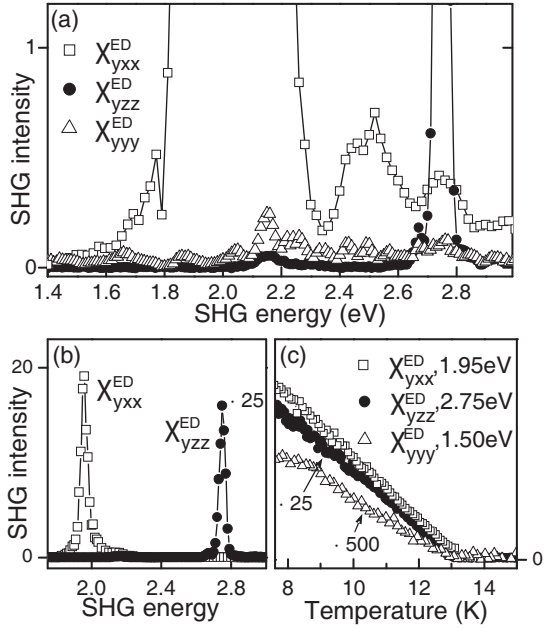}
\caption{\label{fig:fig3} Commensurate SHG in the multiferroic phase of MnWO$_4$ (arbitrary
units). (a, b) C-SHG spectra obtained with $k\,\|\,x$ ($\chi_{yzz}^{\text{\tiny{ED}}}$,
$\chi_{yyy}^{\text{\tiny{ED}}}$) and $k\,\|\,z$ ($\chi_{yxx}^{\text{\tiny{ED}}}$). The exceptionally large
signal yield at 1.95~eV due to multiferroic phase-matching. (c) Temperature dependence of the SHG
intensity from $\chi_{yzz}^{\text{\tiny{ED}}}$, $\chi_{yyy}^{\text{\tiny{ED}}}$, and
$\chi_{yxx}^{\text{\tiny{ED}}}$. All the signals are associated to ED-SHG contributions that are
present in the multiferroic AF2 phase only.}
\end{figure}

Figure~\ref{fig:fig3} shows the SHG spectra on a (100) and a (001) sample in the multiferroic AF2
phase at 8~K. Rich spectra showing all the transitions in Fig.~\ref{fig:fig2}(a) are observed.
We
identify $\chi^{\text{\tiny{ED}}}_{yyy}$, $\chi^{\text{\tiny{ED}}}_{yzz}$, and
$\chi^{\text{\tiny{ED}}}_{yxx}$. According to Table~\ref{tab:chiMnWO} the observed SHG
contributions reflect the change of point symmetry from $2_y/m_y1^{\prime}$ to $2_y1^{\prime}$ by
the emergence of the spontaneous polarization. The assignment is corroborated by temperature
dependent measurements in Fig.~\ref{fig:fig3}(c). The ED-SHG contributions in Fig.~\ref{fig:fig3}
are non-zero only when $P_y^{\rm sp}\neq 0$. All contributions behave according to
$\chi_{ijk}^{\text{\tiny{ED}}}\propto P_y^{\rm sp}$ which, because of $P_y^{\rm
sp}\propto(T_{\text{\tiny{AF2}}}-T)^{1/2}$ and $I_{\rm SHG}\propto|P_y^{\rm sp}|^2$, leads to a
linear temperature dependence of the SHG intensity in the AF2 phase.

We thus see that in spite of the IC magnetic order of MnWO$_4$ the SHG signal can be
fully understood on the basis of the macroscopic point group symmetries thus far. The SHG signal picks up the
magnetically induced spontaneous net polarization as commensurate projection of the IC spin order
while SHG contributions coupling to the underlying IC magnetic structure are not detected. Neither
the polarization-dependent nor the temperature-dependent SHG measurements exhibit peculiarities
revealing the presence of an aperiodic phase.

Another noteworthy issue in Fig.~\ref{fig:fig3}(b) is the exceptionally pronounced resonance of
the $\chi^{\text{\tiny{ED}}}_{yxx}$ component which is due to phase matching.\cite{Boyd08} Nearly
noncritical type I phase matching can be realized in MnWO$_4$ for incidence along $x$ at
1.95~eV, as can be seen from the refractive indices of the crystal.\cite{Becker07a} This constitutes
a rare example of phase matching activated by magnetic order and, because of the associated
spontaneous polarization, presumably the first occurrence ever of ``multiferroic phase matching''.

\subsection{Incommensurate SHG contributions}

\begin{figure}
\centering
\includegraphics[width=0.40\textwidth]{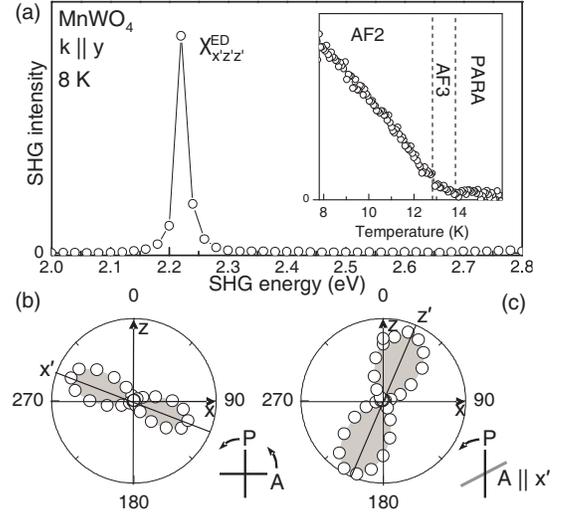}
\caption{\label{fig:fig4}  Incommensurate SHG in the magnetically ordered phases of MnWO$_4$. (a)
IC-SHG spectra obtained with $k\,\|\,y$ for which C-SHG contributions of any multipole order are
not allowed. The temperature dependence in the inset confirms the presence of the IC-SHG signal in
the multiferroic AF3 phase and the non-multiferroic AF2 phase. (b, c) Polarization dependence of
the IC-SHG signal at 8~K and 2.22~eV for light incident parallel to the $y$ axis. $x^{\prime}$ and
$z^{\prime}$ denote the axes of the rotated coordinate system consistent with the polarization
dependence of the IC-SHG signal. P and A denote the polarization of the incident light and of the
detected SHG component, respectively. In (b) P and A are rotated simultaneously while in (c) P is
rotated while A remains constant.}
\end{figure}

Figure~\ref{fig:fig4} shows the SHG spectrum obtained on a (010) sample in the multiferroic AF2
phase at 8~K with $k\,\|\,y$.
A SHG signal from the \mbox{${}^6A_{1g}\rightarrow{}^4T_{1g}$}
transition with a variety of most unusual properties is observed. First, SHG of the ED, MD, and EQ
type is forbidden according to Table~\ref{tab:chiMnWO} if the macroscopic point symmetries
$2_y/m_y1^{\prime}$ or $2_y1^{\prime}$ are applied. Second, the temperature dependent SHG data
reveal that the SHG signal emerges in the non-multiferroic AF3 phase (with a change of slope at
$T_{\text{\tiny{AF2}}}$) in contrast to the SHG signals in Fig.~\ref{fig:fig3}(c). Third, the
polarization dependent data in Figs.~\ref{fig:fig4}(b) and \ref{fig:fig4}(c) reveals lobes
consistent with a coordinate system that is rotated by about $25^{\circ}$ around the $y$ axis of
the Cartesian system. Because of
the violation of the inversion symmetry in the AF2 phase, the observed SHG contribution can be
assigned to an ED process. The corresponding susceptibility is
$\chi_{x^{\prime}z^{\prime}z^{\prime}}^{\text{\tiny{ED}}}$ if the rotated coordinate system ($x^{\prime}$, $y^{\prime}$, $z^{\prime}$) shown
in Fig.~\ref{fig:fig4} is applied. Note that the axes $x^{\prime}$ and $z^{\prime}$ do not only deviate from the $x$ and $y$ axes
but also from magnetic easy or hard axis.\cite{Lautenschlaeger93a}
We speculate that the rotation may be related to the propagation direction perpendicular to the
planes defined by the IC propagation vector $\mathbf k_{\text{\tiny{AF3}}}$. This direction
includes an angle of $28^{\circ}$ with the $x$ axis. All the
aforementioned inconsistencies suggest to associate the ``forbidden'' SHG signal in
Fig.~\ref{fig:fig4} to the IC magnetic order which has not yet been included into the symmetry
analysis. This will be discussed in the following.

Remember that for a symmetry analysis as in Table 1 only {\it global} symmetries are
considered. Site symmetries, defects, and translations, {\it i.e.}, {\it local} properties that
are spatially confined to the extension of the unit cell, are neglected so that only the point group
symmetry of a crystal is considered. For SHG
experiments in the optical range this is a reasonable approximation because the wavelength of the
probe light exceeds the extension of the unit cell by three orders of magnitude. However, IC
phases are characterized by structural modulations {\it exceeding} the extension of the unit
cell. Therefore the spatial variation of the electromagnetic light field across one period of the
IC wave vector may no longer be negligible so that the light begins to sense the {\it local}
symmetry of the IC structure.\cite{Dvorak83a}

Based on the Neumann principle observation of SHG from
$\chi_{x^{\prime}z^{\prime}z^{\prime}}^{\text{\tiny{ED}}}$ indicates the presence of a local
symmetry lower than $2_y1'$. Here, the only matching point group is $11'$. This becomes reasonable
once the local structure of MnWO$_4$ is considered. The spatial variation of the
electromagnetic light field couples to the local violation of the twofold rotation symmetry $2_y$
due to the IC nature of the ellipsoidal magnetic spin wave and leads to the reduction of
the local symmetry from $2_y1'$ to $11'$.

\begin{figure}
\centering
\includegraphics[width=0.45\textwidth]{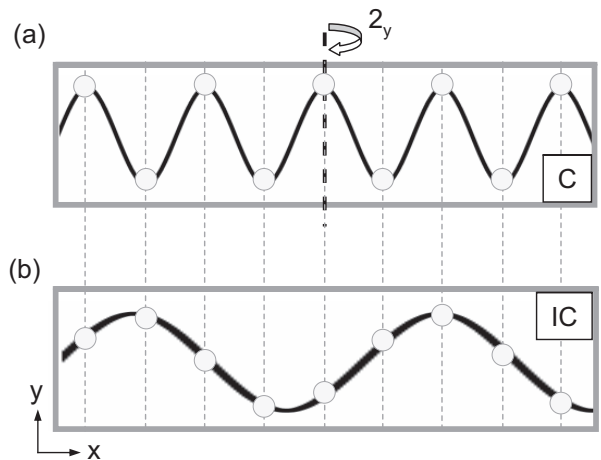}
\caption{\label{fig:fig5} Model explaining the presence of ``forbidden'' SHG contributions in
crystals with IC order. (a) A commensurate modulation of the crystal structure retains the twofold
axis of the structure. (b) An IC modulation violates the twofold symmetry locally. Additional SHG
contributions may therefore emerge.}
\end{figure}

Figure~\ref{fig:fig5} illustrates such a violation of the local symmetry by IC order.
A section
whose extension is determined by the coherence length of the SHG process is considered so that any
SHG light generated in the crystal section contributes coherently to the SHG yield.
Figure~\ref{fig:fig5}(a) shows a commensurably modulated chain of atoms. One can see that the
commensurate modulation conserves the symmetry axis $2_y$ that is indicated by the twofold
rotation axis in the sketch. In contrast, the IC modulation in Figure~\ref{fig:fig5}(b) violates
the lattice periodicity and destroys the rotational symmetry within the considered section. Hence, locally, {\it i.e.}, in the
finite section probed coherently by the SHG process, the $2_y$ symmetry is lost.

\section{Conclusion}

In conclusion, we have analyzed the coupling of optical SHG to an incommensurately ordered state
using multiferroic MnWO$_4$ as model compound. Two principally different SHG contributions
coupling to the IC order parameters were identified. The first one couples to a commensurate
projection of the IC state and is typically represented by a secondary, induced order parameter.
This SHG contribution can be understood on the basis of a point symmetry analysis for commensurate
structures and is indistinguishable from SHG contributions induced by primary commensurate order
obeying the same point symmetry. The second SHG contribution couples to the primary IC order. It
reflects the local symmetry of the crystal on the length scale of the IC modulation. Since the IC
order breaks local symmetries, additional SHG contributions that are forbidden by the global point group
symmetry of the infinitely extended crystal can emerge. They uniquely identify the IC state and
can be used to image IC domains and separate them from any coexisting commensurate domains. Here,
IC domains expand the established concept of ferroic domains because in contrast to commensurate
translation domains neighboring IC domains differ by a lattice translation that exceeds the
boundary of a single unit cell.\cite{Meier09c}

From the point of view of application we have characterized SHG as a tool for studying the
properties of coexisting commensurate and IC forms of order in a single experiment. The local
symmetry reduction in an IC crystal leads directly to the manifestation of additional, ``hidden''
degrees of freedom for forming domains that are accessible by SHG. These additional degrees of
freedom play an important role for the manifestation of the functionalities of materials with IC
order. For instance, in multiferroic MnWO$_4$ the correlation between the magnetically ordered IC
state and its domains on the one hand and the commensurate dielectric properties of the material
on the other hand is a key for understanding the complex magnetoelectric interactions of the
crystal.

\begin{acknowledgments}
This work was supported by the DFG through the SFB608.
\end{acknowledgments}

\newpage


\bibliographystyle{unsrt}

\end{document}